\documentclass{ceab}   

\usepackage{epsfig}     
\usepackage{graphicx}   

\usepackage{ceabbib}     
\usepackage[T1]{fontenc}

\begin{document}

\def\tit{VIII$^{\lowercase{th}}$ HAC Proceedings Template}
\def\aut{Stencel}

\title{Results of the Recent $\epsilon$ Aurigae Eclipse Campaign}

\author{Robert E. Stencel  
\vspace{2mm}\\
\it University of Denver Observatories \\
\it 2930 E. Warren Avenue, Denver Colorado 80210 USA \\
 }

\maketitle

\begin{abstract}
Results of the 2010 eclipse campaign are described, and preliminary interpretations proposed.  These include photometric, interferometric, spectroscopic, astrometric and polarimetric observational results.  Next steps, along with continued monitoring, include simulations and other future work.  Numerous acknowledgements are appropriate for the many participants in making this international effort a success.

\end{abstract}

\keywords{Hvar astrophysical colloquium - proceedings - eclipsing binaries - $\epsilon$ Aurigae}

\section{Introduction}


First, I would like to thank the organizers of this Hvar conference for allowing me to speak about results of the recent campaign to observe $\epsilon$ Aurigae during its 2009--2011 eclipse, and to communicate some of the developments of the past few years.  There is a lot to report, of course, about the nature of the campaign, all the participants as well as the photometric results, interferometric and spectroscopic results, astrometric, polarimetric, lots of different kinds of results.  Plus, a variety of models and simulations that are beginning to develop based on this wealth of data that has been accumulated.  Naturally, no one person did all of this and numerous acknowledgements are appropriate.  A number of reports have been assembled in the online Journal of the American Association of Variable Star Observers, eJAAVSO, volume 40, number 2.

As you know from talks at this meeting by Phil Bennett, Ed Guinan and Petr Harmanec, $\epsilon$ Aurigae is an eclipsing binary system with a 27.1 year period.  Think about the fact that the last eclipse prior to this 2009--2011 event was during the years 1982--1984.  Back then, a large telescope was only a meter or so in aperture.  The Hubble Space Telescope was not yet in orbit.  IUE and IRAS were the new satellites at the time.  Optical-infrared (OIR) interferometry was in its infancy.  Photometry back then involved vacuum tube 1P121 photon counting detectors.  In 1983, there were no CCD cameras, no internet, no cell phones -- lot of technology has changed since then.  

\section{Multi-wavelength Photometry}
Next, let's look at some of the wonderful light curve data compiled during this eclipse-watching campaign.  Hopkins (2012) and AAVSO have compiled UBVRIJH filter photometry, as well as visual reports at the website, www.aavso.org, and there one can see the characteristics of the eclipse through the entire event, 2009 August -- 2011 August, along with the continuing observational contributions to present.  The post-eclipse coverage could be a little better sampled in certain bands, but overall we did better during post-eclipse than some prior eclipse cycles that stopped observing too soon.  Any case, the detail is wonderful.  From the redder photometric bands out to the bluer photometric bands one can see a wavelength dependence of some of the fine pulsation-like detail, and even a hint of differential slopes either side of mid-eclipse (2010 August, JD 2,455,400).   

Thanks to Lou Boyd and Jeff Hopkins, we have nearly continuous (seasonal) UBV  monitoring over the decades, since the 1983 eclipse, to the present.  Kloppenborg, Hopkins and Stencel (2012) have analyzed the irregular out-of-eclipse variation seen in those and additional reports.  Those 0.1 to 0.2 magnitude variations have been noted in the literature for many years, with greater amplitude at shorter wavelengths.  At times, the variations are almost absent: very flat, low-level, erratic-looking variations.  At other times, the variations appear almost periodic.  There is a well-known 67-day periodicity that Kim (2008) and others have discussed.  What Brian Kloppenborg managed to do, as part of his Denver University dissertation, was go through these data and apply a weighted wavelet analysis, in an effort to try to pull out any persistent periodicities.  What he found was that these variations are multi-periodic.  Currently, two periods are prominent, 69 and 88 days, but these apparently are evolving in time, with about a one-and-a-half day per year secular decrease.  The interesting point here is that we have fairly rapid period evolution, which is possibly a sign of rapid interior/envelope evolution.  We noted that every 3,200 days, about one-third of orbital period, the sinusoidal characteristic peaks up.  This leads to a testable prediction - that we will see another stable pulsational behavior approximately during 2014 December.  This provides another reason for observers to continue the photometric coverage of this system.  Indeed, one can ask whether the idea suggested by Saito and Kitamura (1986) has merit, that the F star envelope is shrinking, perhaps due to core expansion, which could abbreviate the period.  Continued interferometric monitoring of the F star diameter might be able to test this (see below).


\section{The Spectral Energy Distribution}
One of the wonderful new things is pan-chromatic coverage from the x-ray region, all the way out to the radio regime.  Hoard, Howell and Stencel (2010) assembled a spectral energy distribution (SED)  based on numerous sources, recently augmented with Herschel Space Observatory sub-mm photometry (Hoard et al. 2012).  The bright F star, of course, dominates the system SED at optical wavelengths by orders of magnitude, in brightness terms.  Now that the eclipsing body has been shown to be disk-like, we identify the infrared excess above the F star photosphere with the disk contribution to the SED, based on ground-based plus Spitzer Space Telescope photometry and spectra, AKARI photometry.  On the ultraviolet side, of course, we see a far-UV excess, as measured by the Hubble Space Telescope spectrographs GHRS, STIS and COS, plus FUSE spectra.  Hoard, Howell and Stencel (2010) ascribed the UV excess as due to a B5V star based on the continuum level and Lyman$\alpha$ line shape.  While that is plausible, one has to factor in possibility of accretion-power contributing to the flux.  Phil Bennett in this meeting provides additional discussion of this.  Overall, the SED defines the relative contributions of the three main actors in this drama: the high-luminosity F star, the disk and a hot star.

Photometrically, we now are able to extend things from the X-rays (Wolk et al. 2010) all the way to nearly the radio regime.  Thanks to the Herschel Space Observatory, my collaborators and I were able to pick up new measurements at 100 and 250 microns (Hoard et al. 2012), and these sub-mm fluxes seem to fit a black-body curve for a 550K cold disk rather well during these recent times.  However, combining this with infrared photometry during the 1990s, we are able to show very clearly that the disk has a warm face (anti-eclipse phase) and a cold face (eclipse phase).  The warm face can be fitted by an 1100K blackbody excess above the F star SED, which occurs when that side of the disk sees the F star.  Noon, if you will.  And the cold face that we see during eclipse is substantially cooler, about 550 Kelvin.  That is a very significant set of evidence for trying to interpret the nature of the disk properties.  We can use these Herschel Space Observatory fluxes to make an estimate of mass of the disk.  The dust mass (Jura et al. 2001) is given by: \\

\indent \indent  M$_{dust}$ = F$_{\nu} \lambda^2$ d$^2$ / ( 2 k T$_{dust} \kappa_{\nu}$  ).       
\\

\noindent Given the observed flux, F$_{\nu}$, 57 mJy at 250 microns, and using the Hipparcos distance (650 pc), the 550K dust temperature for the cold side, and a mass-absorption coefficient (Beckwith et al. 1990),\\

\indent \indent  $\kappa_{\nu}$ = 0.1 ($\nu$/10$^{12}$Hz), 
\\  

\noindent one comes up with a dust mass of 4x10$^{29}$g, or about 0.2 Jupiter masses.  If the disk is transitional, the gas to dust ratio might be well below the interstellar value of 100, suggesting the disk mass is less than a few Jupiter masses in total.  Tabulated F supergiant star mass loss rates of 10$^{-7}$ M$_{\odot} yr^{-1}$ indicate a wind to disk replenishment timescale of {\it only} 60,000 years.  


\section{Interferometric Imaging}

Due to the advancement of optical technology, computers and metrology, we were able to obtain the breakthrough results on $\epsilon$ Aurigae, using the MIRC beam combiner at the CHARA array at Mt. Wilson Observatory.  Until lately, the system has been mysterious, with an invisible companion causing the eclipses.  But, we caught the eclipse-causing body on camera.  We were able to see the ingress of the dark disk itself, and with continued coverage, detect a number of features about the disk: thickness and extent (Kloppenborg, Stencel, Monnier et al. 2010).  Brian Kloppenborg, as part of his Denver University dissertation has modeled the multiple epochs of data throughout the eclipse (Kloppenborg, 2012).  As a result, the eclipse-causing companion is no longer quite as mysterious.  It resembles a generic astrophysical disk, perhaps transitional or debris-like, but we are working on getting those details as well.  In fact, using a combined radial velocity, interferometric image plus astrometry approach (using some of VandeKamp's original data), more precise system parameters are emerging.  Kloppenborg (2012) derives a most likely distance of 737 parsecs, with an uncertainty of 10 percent.  Interferometrically, of course, we have the F star radius in angular units (2.28 milli-arcseconds, Stencel et al. 2008) and then with the newly derived distance, we obtain physical units, with 10 percent uncertainties: 183 solar radii for the F star radius, 7.3 AU for the length of the disk, and 0.6 AU for the thickness of the disk.  These numbers are updates compared to the Nature paper values that some of you might be using.  Brian Kloppenborg's solution seems to imply the high-mass case, q = 1.1.  The mass function then yields a 13.2 and 11.6 solar masses for the components of this system.  However, there are always caveats.  Remember, the disk seen with MIRC is at an infrared H-band wavelength, so we certainly are seeing the optically thickest portions of the disk.  Second, all these dimensions are very sensitive to limb-darkening assumptions.  You know, the F supergiant probably does not share exactly the solar limb-darkening characteristic, and that is not well determined at this point.  Orbit refinements continue to be worked on.  As of this report at Hvar 2012, these are the numbers that we were favoring.

In addition to imaging interferometry, spectral interferometry is possible as well.  Also at the CHARA array, but this time using the VEGA beam combiner, designed by Denis Mourard and colleagues from Observatoire Cote Azur, Nice.  The beauty of this is that we can use strong lines like H$\alpha$, not only to measure the intensity with wavelength, or velocity in this case, but also measure the visibility through the line.  Visibility decreases from a continuum reference value down to a small value, meaning the source is more highly resolved in the core of the H$\alpha$ line.  We measure a phase, a differential offset of light path, of where that signal is relative to the continuum source.  From these data, as reported in Mourard, Harmanec, Stencel et al. (2012) over three years of eclipse time, we see changes in the size and orientation of the F star, in ways that are consistent with the interferometric imaging.  More importantly, spectro-interferometry is giving us evidence about the vertical extent of the atmosphere of the disk itself that could not be seen in the broadband infrared image.  The disk atmosphere appears to vertically extend more than an F star radius.


\section{Spectroscopy}

In addition to interferometric spectroscopy, normal spectroscopy has been pursued in a big way during this eclipse.  Chadima, Harmanec and collaborators (2011) have reported some wonderful monitoring of H$\alpha$ and other lines through the bulk of eclipse.  In fact, what is seen in this set of data includes the fact that the stellar spectrum showed absorption effects even up to three years prior to ingress.  In fact, a major H$\alpha$ absorption event was seen in 2006, not far from periastron.  The silicon and iron lines also in this region of the spectrum showed periodic variations in radial velocities and central intensities.  These are probably affected by the pulsations occurring in and around the F star photosphere.  Once again, spectroscopically, a 67 day period again appears.  The radial velocities and central intensities however, do not seem to vary completely in phase.  That is, the velocity, the central intensity and the other photometry.  They are not all locked together, so they are clearly arising with different line contribution functions, differing atmospheric altitudes.

Additional spectroscopy has been pursued by very capable amateur astronomers.  Robin Leadbeater, notably, used his relatively small telescope and one of the wonderful new LHIRES spectrometers that Olivier Thizy (www.shelyak.com) provided, and was able to monitor the red line of potassium, 7699\AA.  Indeed, he was able to show, following the Lambert and Sawyer (1986) results from the 1985 eclipse, how the line varies strongly from the interstellar level, throughout the eclipse.  The equivalent width grows as a a red absorption component emerges, just as first and second contact come to pass.  Then, through mid-eclipse we see the line centroid move over to the blue side of the line, and finally vanish again in the current year.  Now, in that set of measurements, he agrees rather well with Lambert and Sawyer.  Robin Leadbeater also discovered this interesting quantization of equivalent width with time, through the course of the eclipse.  He and I have interpreted that as indication of sub-structure in the disk (Leadbeater and Stencel, 2010).  The equivalent width pattern begins at a low level and then seems to step up every so often.  It is these ramps that reflect an increase in the amount of material, then a relative lack of change in density, increase, lack, increase, and then a mid-eclipse fading, perhaps due to ionization or other effects on the line strength.  This is followed by a post-mid-eclipse growth, nearly doubling the equivalent width by third contact, and then finally a falloff, beginning close to third contact.  In fact, the equivalent width did not come back down to near zero until early 2012, long after photometric eclipse ended in the UBV bands.   Whether it be spiral arms or rings in the disk remains to be determined, but substructure has been revealed.  Additionally, the rotation of the disk from potassium line radial velocities is a way of getting another estimate of that secondary mass, as well, suggesting a low mass solution.

Additional wonderful spectroscopic monitoring was accomplished by Klaus Straussmeier and friends using their STELLA echelle spectrometer up at the Canary Islands.  Lothar Schanne has been pursuing the data reduction, and has shared a sample of some of the 300 spectra obtained in the interval since the 2006 Prague IAU meeting (when and where I urged Dr. Strassmeier to add $\epsilon$ Aur to his observing plan for STELLA).  The point Lothar makes is that the line shapes follow a variation similar to potassium: we see pre-eclipse oscillations due to the F star 67-day period.  At the start of eclipse, the disk red-shifted material comes into view.  Through mid-eclipse the blue-shifted material appears, and then so on.  Then, it evolves back toward a more normal photospheric behavior at end of the eclipse.  Schanne (private communication)characterized a few hundred of these spectral lines, and finds that about 2/3 show no eclipse effect (typically higher excitation energy), while about 1/3, the lower energy cases, show stronger  effects .

Another of the spectroscopic monitoring efforts is by Elizabeth Griffin at Dominion Astrophysical Observatory, in Canada.  She and I submitted an eJAAVSO paper that talks about new observations using CCD Spectra, plus the fact that she has been continuing to digitize a lot of the classical photographic material from Mt. Wilson and Victoria (Griffin and Stencel, 2012).  What we noted in particular is that a lot of the disk features seem to repeat pretty exactly eclipse to eclipse, over four or five different eclipses, and that means on a human timescale this system is fairly stable in its current configuration.  We note, as have others, that the disk is asymmetric: the second half of the eclipse seems to have more enhanced absorption characteristics than the first half of the eclipse.  Also, right at third contact she found some very narrow absorption features appearing that could well be a mass transfer stream, which is relatively new knowledge about this system, in terms of the literature.  An interesting additional point is that the disk spectrum resembles the chromosphere of zeta Aurigae during partial eclipse.  Comparing the spectrum near third contact, from the prior eclipse, with a recent zeta Aurigae chromospheric eclipse during the 1990s, and they look very similar.  If anything, the disk lines might be slightly narrower, inferring a lower gravity but comparable temperatures and densities to the chromosphere.  Curiously, we also found rare earth lines showing up as slightly enhanced during this third contact period, by comparing an out of eclipse phase with a third contact phase.  The disk contribution is certainly lessening, but these rare earth lines seem to perk up at just certain very narrow range of phases associated with what appears to be a mass transfer stream, perhaps detected during a U and B lightcurve still-stand close to JD 2,455,670 (Hopkins, 2012).

We pursue stellar spectroscopy for lots of reasons, not just to look for interesting changes but to be quantitative about them.  Thankfully there are a couple of groups that have been doing quantitative analysis for abundances, and it is interesting to look at those preliminary results.  For example, Sadakane et al. (2010)  looking at $\epsilon$ Aurigae high-dispersion spectra note that elements like nitrogen, sodium, yttrium and barium seem slightly elevated relative to solar, while carbon and oxygen might be slightly sub-solar.  We are talking maybe 0.3 dex, not a huge effect.  And they refer to this as a pattern similar to other yellow supergiant stars.  Ishigaki et al. (2012) discuss CRL2688, which is a similar type of F supergiant, or perhaps a post-AGB low-mass supergiant.  In that case, he also reports elevated carbon, nitrogen, oxygen, sodium, yttrium, but slightly sub-solar iron and barium.  Barium to yttrium ratio is a proxy for heavy versus light s-process element creation.  $\epsilon$ Aurigae does not seem to be as processed as CRL2688, in those terms.  Here are some more comparisons: carbon to iron, very similar in ratio.  Nitrogen to iron, very comparable.  Sodium to iron, also very similar, within the quoted errors.  So, by itself, an abundance pattern does not eliminate the $\epsilon$ Aurigae primary star from being a post-AGB object.  The low mass scenario deserves further consideration.
  
Another very important set of clues has been seen in ultraviolet spectra, going way back to the GHRS and even IUE times.  Although, not commented upon in those papers, the ionized carbon 1335\AA~lines in our recent spectra using Hubble and COS instruments do seem to show the same pronounced P-Cygni profiles seen much earlier.  With the kind of velocity separation involved therein, 50 to 70 kilometers per second winds can be inferred.  That is a fairly hefty stellar wind speed for an F-type star.  In collaboration with Phil Bennett, we are working up a little report on this topic.  

I am happy to report that we were able to use the NASA IRTF SpeX instrument over the course of several years, to monitor the spectrum of the infrared during eclipse.  To our delight, one of the features that perked up right at mid-eclipse was the helium 10830\AA~line, which has a nearly 20-volt ground state, and so is very high energy, relatively speaking.  In our infrared report, Stencel et al. (2011), we reported the fact that during the course of the eclipse, 10830 was relatively weak, typical of Be stars, and then it peaked up the equivalent width, increasing by an order of magnitude at mid-eclipse, and then slowly fading away over a few months as totality went on.  This seems to be consistent with something like a Stromgren sphere near the middle of the disk, and again with a central object, if it is 5 or 10 solar masses, is going to have a modest amount of UV output ionizing radiation.   During the last eclipse as well, carbon monoxide was noted in the spectrum again after mid-eclipse, during totality.  We made an effort to observe that, and indeed found the same variation, as reported in the aforementioned paper.  However, we have not yet found any other solid-state features.  No actual ices or silicates or carbonate-type mineral bands, and that of course remains a bit of a puzzle.  Now, if the disk is above 500 Kelvin, maybe it is just too warm for those to exist.  

But, in terms of the CO, what does seem to be happening is that prior to mid-eclipse, one sees the near infrared continuum as measured by SpeX to be fading away.  Then, during middle and latter half of eclipse, one sees a step develop right at the carbon monoxide band head at 2.29 microns wavelength.  Fortunately the Gemini 8-meter re-commissioned its GNIRS spectrometer in 2010, and right during eclipse they approved a request for observing time.  We were able to go in and get some very high-resolution data.  From this, one can of course get densities and temperatures and, more importantly, to look carefully at the question of isotopic carbon, 13CO, which is seen longward of 2.34 microns, but which is unfortunately at a wavelength compromised by the Paschen series converging toward its limit.  So, I am in the process of trying to disentangle all those contributions.  The claim during the 1985 eclipse was that isotopic 13 carbon was greatly enhanced compared to solar, and this was a sign that the F star is in a post-AGB evolutionary state.  There probably is 13 carbon present, but whether it is as enhanced as previously claimed, we are hoping that these higher-resolution spectra can help us disentangle that.  Note added in proof: 13 carbon is confirmed in the newer spectra, as is 2.2 micron Na I emission.

There is a lot of additional important spectroscopy that has been done and it has tremendously enriched the archives.  Here is a list of some of the important sources: Dr.~Munari at Asiago, Dr.~Izumiura at Okayama, Dr.~Ketzeback at Apache Point, Dr.~Martin at Northern Illinois, Dr.~Morrison at Toledo, and others, have diligently observed during eclipse and I am urging them all to keep the data accessible and in good archival mode.  It may transpire that at a phase of special interest, perhaps only one observer managed to cover that precise phase.  So, again, these additional data are available and should be utilized.

\section{Polarimetry}

One of the additional facets of this eclipse was a chance to do polarimetry.  For once among the stars, we have a fairly well-defined geometry, thanks to the interferometric images, in terms of how the system changed shape during eclipse.  Polarimetrically during the last eclipse, 1985 timeframe, Jack Kemp and colleagues monitored in UBV, and detected significant polarization changes.  In polarization, 0.5 percent is a major change, as you may know.  In this case, compared to the light curve timing, there was variation during eclipse, probably on the 67-day F~star pulsation kind of timeframe.  Maybe some of the scattering was enhanced by the disk being in the way.  Notably the third contact showed a big jump away from being polarized back toward a more steady state, interstellar light value.  Kemp in his time was able to use this signal, the phase angles and so on, to infer how a disk might have moved across the F~star, and this is remarkably close to what our interferometric imaging discovered.  I am impressed with what polarization studies are capable of doing.  Kemp additionally talked about hot spots on the poles and equatorial rings.  The evidence to confirm is not quite there yet, but there may be some merit to that idea, too.

One of our accomplished amateur astronomers, Gary Cole, was very ambitious, built his own polarimeter, and got out there and obtained these V-band polarimetric measurements expressed in Q and U Stokes parameters.  His record starts at mid-eclipse because of problems getting his instrument working.  What he found actually mirrors very closely what Kemp et al. (1986) were reporting last time around, particularly this third contact reversion back to out of eclipse behavior, and the variations during eclipse.  Cole (2012) reported on this in the online journal, JAAVSO. 

An additional exciting area is the fact that the ESPaDoNS spectro-polarimeter on the Canada-France-Hawaii telescope at Mauna Kea began to monitor this system as well, partly in response to my request to Nadine Manset.  ESPaDoNS is very high-dispersion, high signal to noise data.  Denver graduate student, Kathy Geise has a paper in eJAAVSO (Geise et al. 2012), and is working up more results.  The polarization in the core H$\alpha$ varies over time, due to the F star.  During the eclipse ingress of the disk one sees a strongly red-shifted polarization signal.  During egress, coverage was sparse, but we did see enhanced blue-shifted effects as well, even up to fairly recent days preceding this conference in mid 2012.

\section{Discussion}
What does one do with all of these data?  Basically, try to build a model that is consistent with it, in order to interpret a system like $\epsilon$ Aurigae.  Everyone agrees on the existence of a high-luminosity early F-type star in the system.  Interferometric imaging essentially proves that there is a disk in the system, and any self-respecting disk has to have a central gravitational mass to keep it organized.  Exactly what that is, what kind of central star, is what we are still working on resolving.  Some of the additional parameters we have to work with include the fact that the disk has a high-noon temperature (that portion facing the F star), that is about twice that of the midnight temperature (facing away from the F star).  As the heated material rotates toward the dark night side and begins to cool off, only to repeat the $\sim$3 year cycle when it comes back out on the dawn side.  We will refer to an inner and outer radius for the disk, roughly 1 AU and 4 AU respectively.  Presumably, it gets a little too warm near the central star for dust to exist, and this fact may help us quantify the true maximum luminosity of the central star.   With that schematic in mind, there are various Monte Carlo photon tracking radiative transfer codes available for modeling disks nowadays, and Denver graduate student Richard Pearson has begun to apply these to the study of the disk in $\epsilon$ Aurigae (Pearson and Stencel 2012).  

I have worked with a couple of students to adapt radiative transfer codes to this case, in particular to try to simulate the heating effect of the F~star on the side of the disk that it faces.  If we can get all the parameters pinned down, given that temperature relates to physical distance from a source, this process may be another way of converting the apparent angular sizes and separations to true distances.  As mentioned, Richard Pearson has been trying to model the variation, showing here the effects of accretion rates and stellar mass on how they might affect the observed external disk temperature.  Muthumariappan and Parthasarathy (2012), using a carbonaceous dust formulation, predicted that the disk temperature would be well below observed.  This means that modeling the heating of the disk by the F~star relatively nearby has to be included.  If we are seeking to determine the system distance, and if we know something about the angular sizes, then if we can simulate the disk noon and midnight temperatures we should be able to make our model consistent among all these parameters.  This is not a new idea: Takeuti (1986, 2011) noticed that the disk ought to be differentially heated.  We are just trying to explore the implications, with a little better data to work on.  These factors also allow us to improve the system orbit.  

The latest eclipse has been a wonderful thing to experience, and the campaign has been reasonably successful.  There has been much international cooperation.  The hard work now is to make sense of the data, and make sure to archive it in a useful way.  The observing is never done.  There are reasons to continue observing, for example, given predictions of how the F~star oscillations might vary around the orbit (Kloppenborg, Hopkins and Stencel, 2012).  We also have suggestions that there may be molecules present in the system.  We have proposed SOFIA sub-mm time, and possibly some radio observations would be helpful, using a northern equivalent of ALMA, for example.  In any event, quadrature and apastron are coming along sooner than you would expect.  It might be worth an occasional spectrum or photometric observation of this system, and if it can be put on observing lists, it can be kept on observing lists.  

A very nice facet of the campaign was the opportunity to engage more of the public.  Third-magnitude stars with a fairly infrequent but noticeable eclipse provided an excuse to try to engage new observers.  Through the AAVSO, we created a Citizen Sky website (www.citizensky.org) and campaign to engage new observers in the whole process.  One of the delightful developments was the fact that people soon realized they could use their digital cameras to actually record photometric information and that has been a real plus out of this effort.  A lot of reasonably precise photometry can be accomplished with digital cameras, it turns out.  

Once again, the latest eclipse inspired some new art.  Sketches were converted into actual paintings, even cartoon drawings.  One of our national cartoonists mentioned $\epsilon$ Aurigae in print.  I suspect that is a fairly infrequent occurrence, to have an astronomical theme in a cartoon, especially an obscure astronomical theme.  Some of you may be familiar with Second Life, which was an online 3D reality platform where one could  construct all manner of things.  My collaborator, Jeff Corbin, managed to create a wonderful simulation of $\epsilon$ Aurigae disk and primary star, which was available in Second Life for manipulation.  So I thank him for that, and also the editing of the videotape shared with the Hvar audience in lieu of my physical participation at the meeting.  Finally, again, no individual could have accomplished all this, and many, many parties contributed.  I was privileged to help coordinate some of it, and my thanks to all of them and to you.

%


%
%

\section*{Acknowledgements} 
I am grateful to the organizers of this conference for the invitation to speak, and for support in part from the US National Science Foundation, and the bequest of William Herschel Womble in support of astronomy at the University of Denver.

\section*{References}
\begin{itemize}
\small
\itemsep -2pt
\itemindent -20pt

\item[] Beckwith, S., Sargent, A., Chini, R., Guesten, R., 1990, Astron.J. {\bf 99}, 924. A survey for circumstellar disks around young stellar objects. \\ http://adsabs.harvard.edu/abs/1990AJ.....99..924B .

\item[] Chadima, Harmanec, P., Bennett, P., Kloppenborg, B., Stencel, R., Yang, S., Bozic, H., Slechta, M., Kotkova, L., Wolf, M., Skoda, P., Votruba, V., Hopkins, J., Buil, C. and Sudar, D.  2011 Astron. Astrophys. 530: 146. Spectral and photometric analysis of the eclipsing binary epsilon Aurigae prior to and during the 2009-2011 eclipse. \\ http://adsabs.harvard.edu/abs/2011A\&A...530A.146C .

\item[] Cole, G. 2012 JAAVSO {\bf 40}, No. 2, 787. Polarimetry of epsilon Aurigae, From November 2009 to January 2012. \\ http://adsabs.harvard.edu/abs/2012JAVSO..40..787C .

\item[] Geise, K., Stencel, R. E., Manset, N., Harrington, D. and Kuhn, J. 2012 JAAVSO {\bf 40}, No. 2, 767. Eclipse Spectropolarimetry of the $\epsilon$ Aurigae System. \\ http://adsabs.harvard.edu/abs/2012JAVSO..40..767G .

\item[] Griffin, R.E.M. and Stencel, R.E. 2012 J.AAVSO {\bf 40}, No. 2, 714.  UV-Blue CCD and Historic Photographic Spectra of epsilon Aurigae.  \\ http://adsabs.harvard.edu/abs/2012JAVSO..40..714G . 

\item[] Hoard, D., Howell, S., Stencel, R., 2010 Astrophys.J. {\bf 714}, 549.  Taming the invisible monster: system parameters for $\epsilon$ Aurigae from the far UV to the mid-IR.  http://arxiv.org/abs/1003.3694 .

\item[] Hoard, D., Ladjal, D., Stencel, R., Howell, S. 2012 Astrophys.J. {\bf 748}, 28.  The Invisible Monster Has Two Faces: Observations of $\epsilon$ Aurigae with the Herschel Space Observatory. http://arxiv.org/abs/1202.6643 .

\item[] Hopkins, J. 2012 J.AAVSO {\bf 40}, No. 2, 633. The International epsilon Aurigae Campaign 2009 Photometry Report. \\ http://adsabs.harvard.edu/abs/2012JAVSO..40..633H .

\item[] Ishigaki, M., Parthasarathy, M., Reddy, B., Garcia-Lario, P.,  Takeda, Y., Aoki, W., Garcia-Hernandez, D. and Manchado, A. 2012 MNRAS {\bf 425}, 997. The Chemical composition of the post-AGB F-supergiant CRL 2688. \\ http://arxiv.org/abs/1205.6101 .

\item[] Jura, M., Webb, R., Kahane, C. 2001, Astrophys.J. {\bf 550}, 71. Large Circumbinary Dust Grains around Evolved Giants? \\ http://adsabs.harvard.edu/abs/2001ApJ...550L..71J .

\item[] Kemp, J., Henson, G.D., Kraus, D., Beardsley, I., Carroll, L., Ake, T., Simon, T. and Collins, G., 1986 Astrophys. J. {\bf 300}, L11.  Epsilon Aurigae - Polarization, light curves, and geometry of the 1982-1984 eclipse. \\ http://adsabs.harvard.edu/abs/1986ApJ...300L..11K .
 
\item[] Kim, H. 2008 Korean Journal of Astronomy and Space Sciences. {\bf 25}, 1- Period Analysis for the F Component of the $\epsilon$ Aurigae System Using Wavelets. http://adsabs.harvard.edu/abs/2008JASS...25....1K .

\item[] Kloppenborg, B., Stencel, R., Monnier, J.D., Schaefer, G., Zhao, M., Baron, F., McAlister, H. A., ten Brummelaar, T. A., Che, X., Farrington, C.D., Pedretti, E., Sallave-Goldfinger, P.J., Sturmann, J., Sturmann, L., Thureau, N., Turner, N., and Carroll, S., 2010 Nature (Letters) {\bf 464}, 870.  Infrared images of the transiting disk in the $\epsilon$ Aurigae system. http://arxiv.org/abs/1004.2464.
 
\item[] Kloppenborg, B. 2012 Dissertation, University of Denver, June 2012.

\item[] Kloppenborg, B., Hopkins, J. and Stencel, R., 2012, JAAVSO vol. {\bf 40}, No.2, 647. An analysis of the long-term photometric behavior of $\epsilon$ Aurigae. \\ http://adsabs.harvard.edu/abs/2012JAVSO..40..647K .

\item[] Lambert, D. and Sawyer, S. 1986 PASP 98: 389 - Epsilon Aurigae in eclipse. II - Optical absorption lines from the secondary. \\ http://adsabs.harvard.edu/abs/1986PASP...98..389L .

\item[] Leadbeater, R. and Stencel, R. 2010, http://arxiv.org/abs/1003.3617 . Structure in the disc of epsilon Aurigae: evidence from spectroscopic monitoring of the neutral potassium line during eclipse ingress.

\item[] Mourard, D., Harmanec, P., Stencel, R., Chesneau, O., Nardetto, N., Perraut, K., Tallon-Bosc, K.,  Berio, Ph., Ligi, Ph. Stee, H. McAlister, T. ten Brummelaar, S. Ridgway, J. Sturmann, L. Sturmann, N. Turner, C. Farrington and P.J. Goldfinger.  2012 Astron. Astrophys. {\bf 544}, 91. High angular and spectral resolution views on the complex system of $\epsilon$ Aurigae . \\http://adsabs.harvard.edu/abs/2012A\&A...544A..91M

\item[] Muthumariappan, C. and Parthasarathy, M. 2012 MNRAS {\bf 423}, 2075. Two-dimensional Monte Carlo radiative transfer modelling of the disc-shaped secondary of Epsilon Aurigae. \\ http://adsabs.harvard.edu/abs/2012MNRAS.423.2075M .

\item[] Pearson, R.L. and Stencel, R.E. 2012 J.AAVSO {\bf 40}, No. 2, 802. Modeling the Disk in the $\epsilon$ Aurigae System: a Brief Review With Proposed Numerical Solutions. http://adsabs.harvard.edu/abs/2012JAVSO..40..802P . 

\item[] Sadakane, K., Kambe, E., Sato, B., Honda, S. and Hashimoto, O. - Publ. Astron. Soc. Japan 62, pp.1381-1390 (2010) - An Abundance Analysis of the Primary Star of the Peculiar Eclipsing Binary $\epsilon$ Aurigae out of the Eclipsing Phase - http://pasj.asj.or.jp/v62/n6/620612/620612.pdf .

\item[] Saito, M. and Kitamura, M. 1986 Astrophys. Space Sci. {\bf 122}, 387. Possible shrinking of the primary component of $\epsilon$ Aurigae. \\ http://adsabs.harvard.edu/abs/1986Ap\&SS.122..387S .

\item[] Stencel, R. E., Creech-Eakman, M., Hart, A., Hopkins, J. L., Kloppenborg, B. K., \& Mais, D. E. 2008, Astrophys. J., {\bf 689}, L137.  Interferometric Studies of the extreme binary, $\epsilon$ Aurigae: Pre-eclipse Observations.  \\ http://arxiv.org/abs/0810.5382 . 

\item[] Stencel, R., Kloppenborg, B., Wall, R., Hopkins, J., Howell, S., Hoard, D., Rayner, J., Bus, S.,  Tokunaga, A., Sitko, M., Bradford, S., Russell, R., Lynch, D., Hammel, H., Whitney, B., Orton, G., Yanamandra-Fisher, P., Hora, J., Hinz, P., Hoffmann, W. and Skemer, A. 2011 Astron. J. {\bf 142}, 174.  Infrared Studies of $\epsilon$ Aurigae in eclipse. \\ http://adsabs.harvard.edu/abs/2011AJ....142..174S .

\item[] Takeuti, M. 1986 Astrophys. Space Sci. {\bf 121}, 127.  An accretion disc surrounding a component of Epsilon Aurigae.  \\ http://adsabs.harvard.edu/abs/1986Ap\&SS.121..127T .

\item[] Takeuti, M. 2011 PASJ: Publ. Astron. Soc. Japan {\bf 63}, 325. Effect of Irradiation on the Disk of the Epsilon Aurigae System. \\ http://adsabs.harvard.edu/abs/2011PASJ...63..325T .

\item[] Wolk, S., Pillitteri, I., Guinan, E. and Stencel, R. 2010 Astron.J. {\bf 140}, 595.  XMM-Newton Observations of the Enigmatic Long Period Eclipsing Binary $\epsilon$ Aurigae: Constraining the Physical Models. \\ http://adsabs.harvard.edu/abs/2010AJ....140..595W .

\end{itemize}

\end{document}